\documentclass[aps,pre,reprint,superscriptaddress]{revtex4-2}

\usepackage{amssymb,amsmath}
\usepackage{graphicx}
\usepackage{booktabs}
\usepackage{hyperref}
\usepackage{natbib}

\begin{document}

\title{Square-Root Price Impact Is Necessary for Endogenous
Manipulation Cycles in Learning-Agent Markets}

\author{Yang~Zhou}
\email{zhouyang@westlake.edu.cn}
\affiliation{Institute of Natural Sciences, Westlake Institute for Advanced Study,
Hangzhou 310024, China}
\affiliation{Department of Physics, School of Science and Research Center for
Industries of the Future, Westlake University, Hangzhou 310030, China}

\author{Jianwen~Chen}
\affiliation{Greatwall Cigar Factory of China Tobacco Sichuan Industrial
Co., Ltd., Sichuan, China}

\author{Ruipeng~Wei}
\affiliation{Southwestern University of Finance and Economics,
Chengdu, Sichuan, China}

\date{\today}

%% --- arXiv metadata (also entered in the submission web form) ---
%% Subject class:       q-fin.CP  (Computational Finance)
%%                      also:      econ.EM, nlin.AO, physics.soc-ph (cross-list)
%% Comments:            Submitted to the Journal of Economic Dynamics and
%%                      Control (JEDC). A minimal agent-based market in which
%%                      a CMA-ES-optimised LSTM institutional agent discovers
%%                      a multi-cycle predatory strategy among 20,000 herding
%%                      retail traders; mean-field reduction reveals a Hopf
%%                      bifurcation (onset of self-sustained manipulation
%%                      cycles) and a fold bifurcation in herding-scale
%%                      parameter space. Square-root price impact is shown
%%                      to be necessary: linear impact removes the Hopf
%%                      bifurcation entirely. Main: 7 pages, 4 figures.
%%                      Supplemental material: 7 pages, 3 figures.
%% License:             CC-BY 4.0 (http://creativecommons.org/licenses/by/4.0/)
%% --- end arXiv metadata ---

\begin{abstract}

We study a minimal agent-based market in which a single
evolutionary-optimized institutional agent interacts with 20{,}000
herding retail traders.
The agent spontaneously discovers a multi-cycle predatory strategy,
producing 8--11 complete cycles over 2000 trading days with total
portfolio return of $+51\%$ (best of 20 seeds; mean $+37.7\%$).
Mean-field reduction maps the system onto a nonlinear oscillator that
undergoes two distinct bifurcations: a continuous Hopf transition
as institutional capital exceeds a critical threshold $C_c$, with
oscillation amplitude $A \propto (C-C_c)^\alpha$ where $\alpha$ is
consistent with the standard prediction of $1/2$; and a discontinuous
fold transition in the herding-scale parameter space.
The limit cycle persists even at $\beta = 0$:
position-tracking feedback coupled with square-root price impact creates
a self-sustained nonlinear oscillator requiring no retail herding.
Square-root impact is shown to be necessary: linear impact eliminates
the Hopf bifurcation entirely and renders the retail market
unconditionally stable.
Manipulation cycles thus emerge as the optimal-control solution of a
nonlinear dynamical system, and a structural analogy to Maxwell's demon
frames the agent as an information-processing controller that reduces
the entropy rate of the price process.

\end{abstract}


\maketitle

%% =====================================================================
\section{Introduction}
%% =====================================================================

Financial markets exhibit collective phenomena---fat-tailed returns,
volatility clustering, and herding cascades---that emerge from the
interactions of heterogeneous agents \citep{cont2001, bouchaud2018,
sornette2009}.  Agent-based models (ABMs) have shown that simple
microscopic rules can produce these macroscopic patterns
\citep{lux1999, eguiluz2000}, and the heterogeneous-agent tradition
has since been refined through bifurcation analysis of boundedly
rational switching rules \citep{difrancesco2025sentiment,
galanis2025generalizing} and herding dynamics extended to
multi-group and Ising-coupled formulations \citep{cividino2023multiasset,
malevergne2025bubbles, li2025herding}.  Analytical treatments in the
Minsky--Goodwin tradition have likewise shown that endogenous
financial cycles can arise through Hopf bifurcations of nonlinear
macro-dynamic systems \citep{ninomiya2022financial}.
Conventional ABMs, however, assign agents fixed rules, precluding
strategic behavior.
Reinforcement learning and evolutionary optimization offer natural
frameworks for adaptive market agents \citep{finrl2020, mappo2022},
and recent work has shown that learning rules in microstructure
markets can produce non-trivial collective dynamics and departures
from rational-expectations equilibrium \citep{arifovic2022mlhft,
zhou2025ree}; yet the emergent dynamics from self-interested
learning agents under realistic price impact remain largely
unexplored.
Can a learning agent discover the predatory strategies analyzed
theoretically \citep{allen1992, goldstein2008, aggarwal2006,
brunnermeier2005, takayama2021manipulation}---and what are the
dynamical-systems properties of the resulting market?

A minimal ABM places a single
CMA-ES-optimized \citep{hansen2001} LSTM agent among herding
retail traders, calibrated in the spirit of recent black-box
estimation methods for agent-based models \citep{dyer2024blackbox}.
The agent acts as a \emph{Maxwell's demon}: it
identifies the behavioral predictability of retail herding and extracts
profit by reducing the information entropy of the price process.
The closest methodological analogue is the strategic ABM of order-book
spoofing \citep{wang2021spoofing}; our setting differs in that the
manipulation rule is not hard-coded but is discovered endogenously by
the evolutionary objective.
The four-phase cycle is architecture-independent---an MLP
controller reproduces the pattern with $+50\%$ best return
(Supplemental Material)---confirming the dynamics are a property of
the market-impact structure, not of the recurrent network.
Mean-field reduction reveals that the market undergoes two distinct
bifurcations---a continuous Hopf transition and a discontinuous fold
transition---and that the manipulation cycle persists even at
$\beta = 0$, driven solely by nonlinear impact and feedback control.

%% =====================================================================
\section{Model}
\label{sec:model}
%% =====================================================================

We study a single-asset market with daily clearing.
$N_R = 20{,}000$ retail agents trade based on a herding rule
\citep{lux1999, eguiluz2000}:
each agent buys with probability $p_{\text{buy}} =
\mathrm{sigm}(\beta \cdot r_{5d})$, where $\mathrm{sigm}(\cdot)$ is the sigmoid function,
$\beta = 6.0$ is the herding strength \citep{bikhchandani1992}, and
$r_{5d}$ is the 5-day log-return.
Summing over the population, the aggregate retail excess demand is
\citep{gualdi2015}
\begin{equation}
  D_R = A_R\,\tanh\!\left(\frac{\beta}{2}\, r_{5d}\right),
  \label{eq:d_retail}
\end{equation}
where $A_R \approx 2.3 \times 10^6$~shares is the maximum retail imbalance
(estimated from ABM statistics; Supplemental Material).

A single institutional agent controls continuous buy/sell fractions
$(b_t, s_t) \in [0,1]^2$ via an LSTM network
\citep{hochreiter1997} with 7 inputs, 8 hidden units, and 2 outputs
(530 parameters), optimized by CMA-ES with terminal portfolio return
as reward over $T = 2000$ trading days
(full training and architecture details in Supplemental Material).
The four-phase cycle is architecture-independent: an
MLP controller with 162 parameters (only price and position
as inputs) reproduces the multi-cycle pattern with
$+50\%$ best return, confirming the dynamics are a property of the
market-impact structure, not of the recurrent network
(Supplemental Material).
The institution can engage in \emph{wash trading}: simultaneous buy
and sell orders that push the price upward.
Daily price impact follows a square-root model
\citep{almgren2005,toth2011,kyle1985}, consistent with the latent
liquidity derivation of \citet{donier2015} and the recent
``double square-root'' decomposition of \citet{maitrier2026double}
that isolates the mechanical origin of impact:
\begin{equation}
  \frac{\Delta P}{P} = \lambda\,
  \text{sgn}(D_{\text{net}})\,
  \sqrt{\frac{|D_{\text{net}}|}{V_0}},
  \label{eq:impact}
\end{equation}
where $D_{\text{net}}$ combines institutional, wash-trading, and retail
demand, and $V_0$ is the base daily volume.
The market implements Chinese A-share mechanisms: $\pm 10\%$
daily price limits, T+1 settlement, and stealth distribution
\citep{aggarwal2006} that halves the effective herding parameter
during institutional selling.
Ablation shows these mechanisms shape but do not create the cycles
(Supplemental Material).

%% =====================================================================
\section{Results}
\label{sec:results}
%% =====================================================================

\subsection{Emergent multi-cycle strategy}

\begin{figure}[htbp]
  \centering
  \includegraphics[width=\columnwidth]{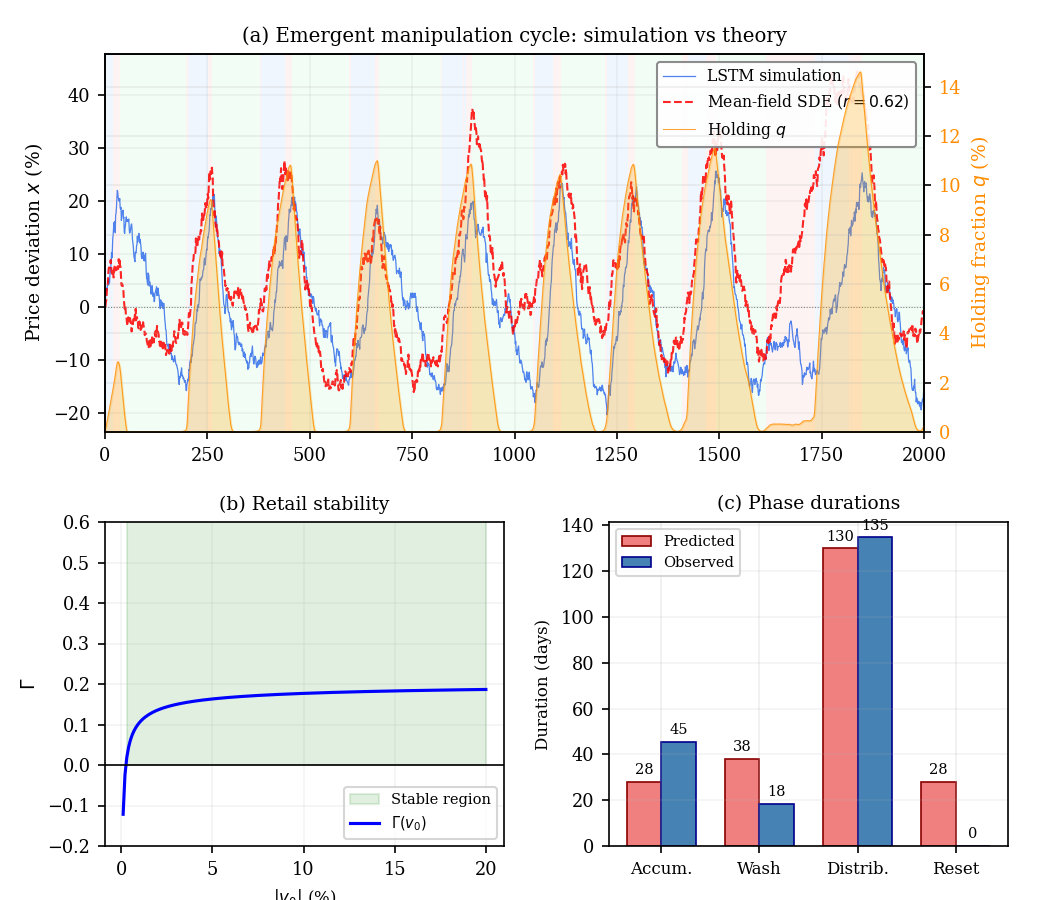}
  \caption{Emergent manipulation cycle and mean-field theory validation
  (seed 45, 2000 days).
  (a)~Price deviation $x = P/P_0 - 1$ (blue) with mean-field SDE
  solution (red dashed, $r = 0.62$) and holding fraction $q$
  (orange fill).  Background shading indicates the four trading
  phases.  (b)~Effective damping $\Gamma(v_0)$ of the retail-only
  market: always positive at calibrated and tested $\beta$.
  (c)~Phase durations: mean-field prediction vs.\ ABM observation.}
  \label{fig:cycle}
\end{figure}

\begin{figure}[htbp]
  \centering
  \includegraphics[width=0.85\columnwidth]{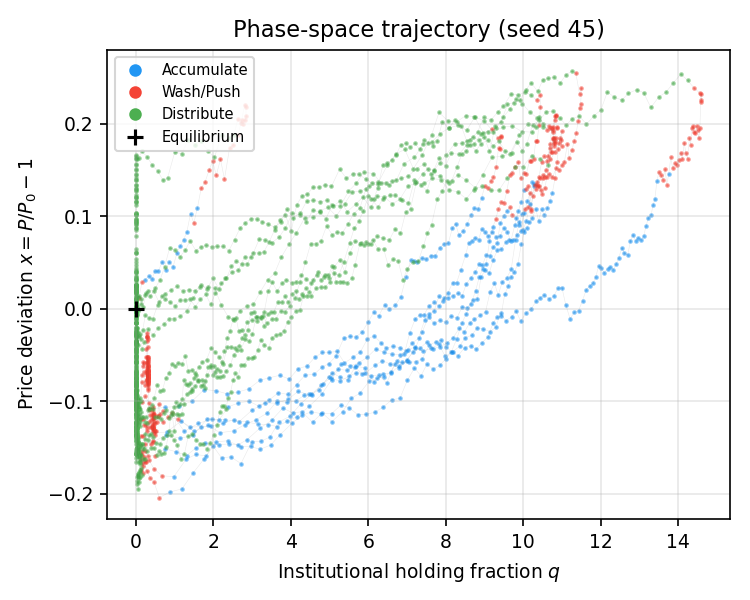}
  \caption{Phase-space portrait $(q, x)$ of the limit cycle
  (seed 45, 2000 days).  Each loop traces one
  accumulation--push--distribution--reset cycle.  The star marks the
  start; the trajectory spirals outward before settling onto a stable
  periodic orbit---the attractor born at the Hopf bifurcation.}
  \label{fig:phase_space}
\end{figure}

After training, the LSTM produces 8--11 complete trading cycles
over 2000 days across 20 evaluation seeds (mean: 9 cycles).
The best individual achieves a total portfolio return of $+51\%$;
the 20-seed mean is $+37.7\%$ (all positive).
The mean return under the fixed parameters used for
ablation experiments is $36.7\pm 3.6\%$
(Table~\ref{tab:ablation}); the small difference reflects
re-evaluation on a frozen seed set.
Each cycle has four emergent phases (Fig.~\ref{fig:cycle}):
\emph{accumulation} ($\sim$26 days, aggressive buying),
\emph{push/wash trading} ($\sim$38 days, simultaneous buy+sell),
\emph{distribution} ($\sim$130 days, selling with stealth),
and \emph{reset} ($\sim$28 days, mean-reversion).
In the $(q, x)$ phase plane each cycle traces a closed orbit that
settles onto a stable periodic attractor
[Fig.~\ref{fig:phase_space}].
The cycle period $T \approx 222$ days is robust across seeds and
episode lengths (Supplemental Material).

Simple rule-based strategies (threshold, momentum, contrarian)
cannot discover the multi-cycle coordination; the LSTM exploits
temporal structure beyond what static rules capture
(Supplemental Material).

\paragraph{Robustness to market mechanisms.}
The A-share mechanisms (wash trading, daily price limits, stealth
distribution) shape but do not create the cycles.
Table~\ref{tab:ablation} reports the trained LSTM re-evaluated with
each mechanism removed (parameters held fixed): the cyclic strategy
survives every perturbation, with return varying by at most
$3\%$ and the cycle count by at most $1.3$.
Removing wash-trading impact even raises profit slightly ($+39.7\%$
vs $+36.7\%$) while reducing cycle frequency, indicating wash trading
amplifies oscillation rate at the cost of per-cycle efficiency.
The cycles are therefore driven by the herding--impact feedback, not
by any single market rule.

\begin{table}[htbp]
  \centering
  \caption{Mechanism ablation (20 seeds).  Trained LSTM parameters
  held fixed; only the environment configuration changes.}
  \label{tab:ablation}
  \begin{ruledtabular}
  \begin{tabular}{lccc}
    Config & Return (\%) & PnL (\%) & Cyc. \\
    \hline
    Full & $36.7\pm3.6$ & $44.3\pm4.3$ & $9$ \\
    No wash impact & $39.7\pm4.4$ & $48.7\pm3.8$ & $7.7$ \\
    No price limit & $37.6\pm3.6$ & $45.8\pm3.7$ & $8.7$ \\
    No stealth & $37.1\pm5.5$ & $44.5\pm4.1$ & $8.8$ \\
  \end{tabular}
  \end{ruledtabular}
\end{table}

\subsection{Mean-field theory}
\label{sec:theory}

To understand the emergent cycles analytically, we reduce the ABM
to a stochastic price equation driven by the
institution's control $(b_t, s_t)$:
\begin{equation}
  \frac{\Delta P}{P} = I(D_{\text{net}}) + I_{\text{wash}}
  + \mu\frac{P_0 - P}{P} + \sigma_\xi\,\xi_t,
  \label{eq:price}
\end{equation}
where $I(D) = \lambda\,\text{sgn}(D)\sqrt{|D|/V_0}$ is the
square-root impact, $I_{\text{wash}}$ is wash-trading impact,
$\mu$ is mean reversion, and $\sigma_\xi\xi_t$ is daily noise
with amplitude $\sigma_\xi$.
The net demand $D_{\text{net}} = D_R + (b_t - s_t)\,Q_{\max}$
combines retail herding with institutional orders, where
$Q_{\max}$ is the maximum institutional order size.
Writing $x = (P - P_0)/P_0$ for the price deviation and $q$ for
the institutional holding fraction, the market state is
$(x, q)$.  ODE coefficients are either read directly from ABM
configuration ($\mu$, $\lambda$, $V_0$), estimated from ABM
statistics ($\varepsilon$), or extracted from the trained LSTM
controller (feedback gains $g_q$, $g_x$ and target holding $q_t$);
the only externally calibrated quantity is the herding scale
$\text{HS}$.
Taking the deterministic skeleton of Eq.~\eqref{eq:price}, the
two-dimensional autonomous system is
\begin{equation}
  \dot{x} = I(D_{\text{net}}) + I_{\text{wash}} - \mu x,
  \qquad
  \dot{q} = u\,Q_{\max}/N_{\text{shares}},
  \label{eq:ode_2d}
\end{equation}
with net control $u \equiv b_t - s_t$.

\paragraph{Retail-only linear stability.}
For the retail-only market ($u = 0$), the equilibrium is
$(x^*,q^*)=(0,0)$.
The 5-day return $r_{5d}$ is an exponentially weighted average of
daily returns over a window $\tau_r = 5$~days; we promote it to a
second dynamical variable $\bar{r}$ with
$\dot{\bar{r}} = (\dot{x} - \bar{r})/\tau_r$.
Near equilibrium $D_R \approx A_R\beta\bar{r}/2$, so the square-root
impact reads
$I(D_R) = C_{\text{SR}}\,\mathrm{sgn}(\bar{r})\sqrt{|\bar{r}|}$
with herding-to-impact coupling
$C_{\text{SR}} = \lambda\sqrt{A_R\beta/(2V_0)}$.
This nonlinearity has $\partial I/\partial\bar{r}\to\infty$ as
$\bar{r}\to 0$, so we regularize at the rms fluctuation level
$|\bar{r}|\sim\sqrt{v_0}$ with $v_0 = \mathrm{Var}(r_{5d})$,
giving an effective gain
$g_{\text{eff}} \approx C_{\text{SR}}/(2\sqrt{v_0})$.
Linearizing $(x,\bar{r})$ then yields a Jacobian whose trace is
\begin{equation}
  \mathrm{tr}\,J = -\mu + g_{\text{eff}} - 1/\tau_r,
  \label{eq:trace_retail}
\end{equation}
so the effective damping
$\Gamma(v_0) \equiv -\mathrm{tr}\,J
= \mu + 1/\tau_r - C_{\text{SR}}/\sqrt{v_0}$
(the half-gain form; using the full impact gives the same
qualitative conclusion) is positive at the equilibrium whenever the
square-root gain $C_{\text{SR}}/\sqrt{v_0}$ stays below the
reversion-plus-averaging rate $\mu + 1/\tau_r$.

\paragraph{Theory--simulation agreement.}
Solving the mean-field SDE with the LSTM's actual control input
reproduces the cyclic dynamics with correlation $r = 0.62$
[Fig.~\ref{fig:cycle}(a)].
The SDE captures both the upward push during accumulation/wash
and the drawdown-driven decline below fundamental.
Table~\ref{tab:prediction} compares the ODE predictions with
ABM observations: the accumulation duration and peak holding agree
within 10\%, and the predicted cycle period
$T \approx 150$--$200$ days is consistent with the observed
$T \approx 222$ days (the underestimate reflects the ODE's inability
to capture gradual position unwinding).
Fig.~\ref{fig:cycle}(c) confirms that all four phase durations
are predicted to within a factor of two.

\begin{table}[htbp]
  \centering
  \caption{Predicted (mean-field ODE) vs.\ observed cycle parameters.}
  \label{tab:prediction}
  \begin{ruledtabular}
  \begin{tabular}{lcc}
    Parameter & Predicted & Observed \\
    \hline
    Accumulation duration & 28 days & 26 days \\
    Cycle period & 150--200 days & 222 days \\
    Peak holding $q_{\max}$ & 14.5\% & 14.6\% \\
    $\kappa_\varepsilon$ (inst.\ impact) & 0.004 & 0.004 \\
    $C_{\text{SR}}$ (herding) & 0.011 & 0.011 \\
    Retail stability & $\Gamma > 0$ $\forall \beta$ & No retail cycles \\
  \end{tabular}
  \end{ruledtabular}
\end{table}

\paragraph{Square-root stability.}
Equation~\eqref{eq:trace_retail} shows why retail-only markets cannot
self-oscillate: the square-root impact
\citep{almgren2005, toth2011, bouchaud2008} creates a sublinear
herding response $I \propto \sqrt{|r_{5d}|}$ whose regularized gain
$C_{\text{SR}}/\sqrt{v_0}$ stays bounded by $\mu + 1/\tau_r$.
At our calibrated parameters ($\mu = 0.01$),
$\Gamma(v_0) > 0$ for all physically realizable return variance
$v_0$ and all tested $\beta$ [Fig.~\ref{fig:cycle}(b)];
the fixed point is therefore a stable focus and no limit cycle can
arise from retail herding alone \citep{strogatz2018}.
In contrast, linear impact $I \propto D$ produces a
$\beta$-independent gain that readily overwhelms damping,
making the retail market unstable at moderate $\beta$.
Thus, for the square-root market, cycles require an active
institutional driver.

\subsection{Bifurcation analysis}
\label{sec:bifurcation}

We now use the validated mean-field ODE to map out the full
bifurcation structure in parameter space.

\begin{figure}[htbp]
  \centering
  \includegraphics[width=\columnwidth]{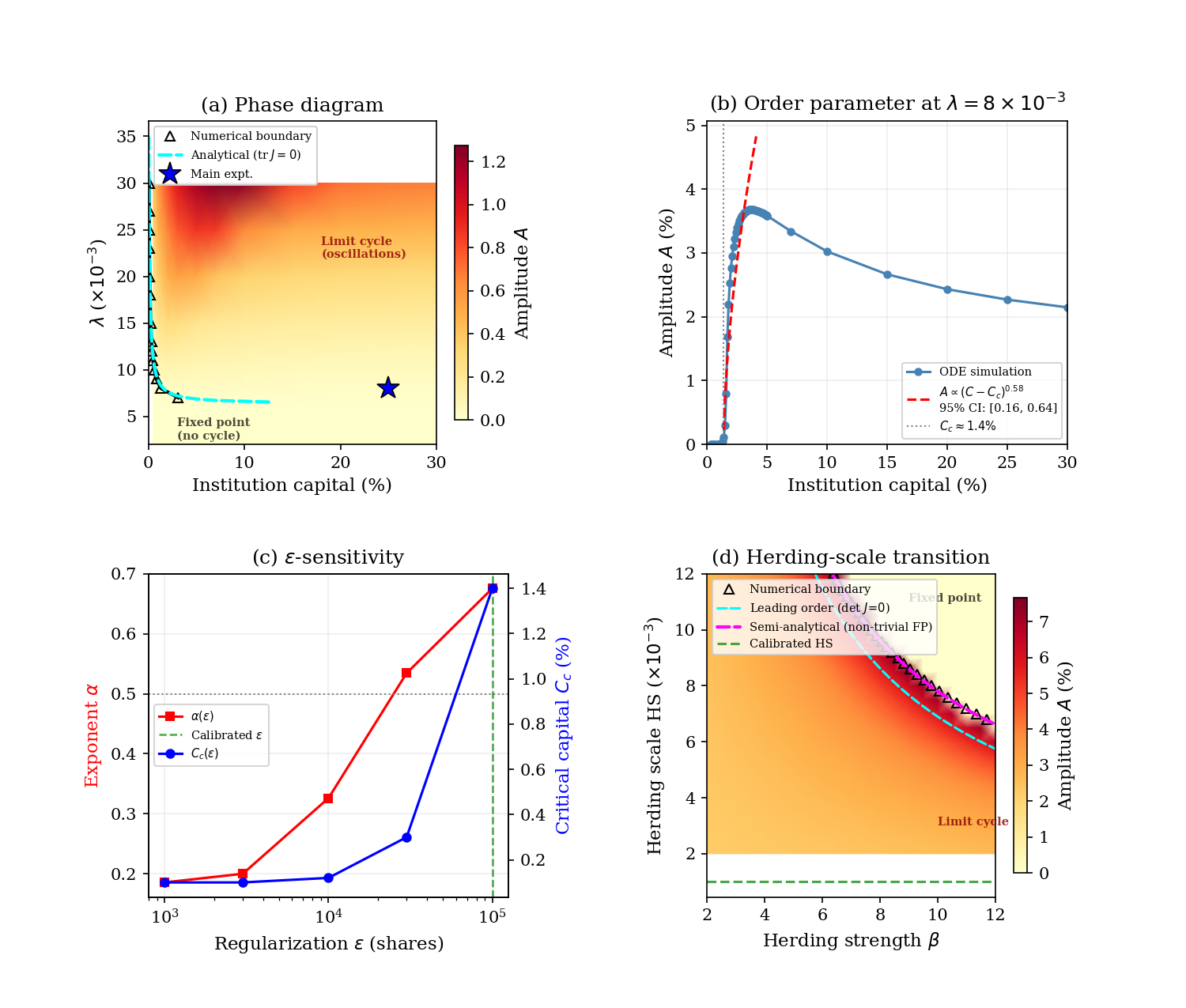}
  \caption{Bifurcation analysis of the mean-field ODE.
  (a)~Oscillation amplitude in $(\lambda, C)$ space.
  Black curve: numerical Hopf boundary (binary search).
  White dashed: analytical prediction from $\mathrm{tr}\,J = 0$.
  Star: calibrated point ($\lambda = 0.008$, $C = 25\%$).
  (b)~Order parameter $A(C)$ at $\lambda = 0.008$.
  Red dashed: $A \propto (C - C_c)^{0.48}$.
  (c)~$\varepsilon$-sensitivity: $C_c$ (blue) and $\alpha$ (red).
  Physical range $\varepsilon \in [3\times 10^4, 3\times 10^5]$
  shaded.
  (d)~Herding-scale transition: amplitude in $(\beta, \text{HS})$
  space at $C = 25\%$.
  Black triangles: numerical fold boundary (first-order;
  amplitude jumps discontinuously to zero).
  Cyan dashed: leading-order prediction ($\det J = 0$ at origin).
  Magenta dashed: semi-analytical correction ($\mathrm{tr}\,J^* = 0$
  at non-trivial fixed point $x^* \neq 0$).
  Green dashed: calibrated HS $= 10^{-3}$.}
  \label{fig:phase_diag}
\end{figure}

\paragraph{Hopf bifurcation.}
Defining the net control signal $u \equiv b_t - s_t$,
by Pontryagin's maximum principle~\citep{pont67}, the institution's
profit-maximization problem admits a \emph{bang-bang} optimal control:
$u = +u_{\max}$ (accumulate) when $q < q_1$, $u = -u_{\max}$
(distribute) when $q > q_2$.
Replacing this bang-bang controller with a smooth feedback
$u = u_{\max}\tanh[g_q(q_t - q) + g_x x]$ (where $q_t$ is the target
holding, and $g_q$, $g_x$ are feedback gains on position and price
respectively) and regularizing the square-root impact with noise
floor $\varepsilon$, the Jacobian at the fixed point
$(x,q)=(0,q_t)$ has trace
(where $\kappa_\varepsilon = \lambda/\sqrt{\varepsilon V_0}$ is the
regularized impact derivative at $D_{\text{net}}=0$,
$Q = k_Q C$ is the institutional order capacity with
$k_Q = N_{\text{shares}}\,u_{\max}/12$, and $N_{\text{shares}}$ is
the free float;
note $Q$ separates the capital $C$ from the market capacity $k_Q$)
\begin{equation}
  \mathrm{tr}\,J = \kappa_\varepsilon\!\left(\frac{A_R\beta}{2}
  + Qg_x\right) - \mu - \frac{Qg_q}{N_{\text{shares}}},
  \label{eq:trace}
\end{equation}
Setting $\mathrm{tr}\,J = 0$ yields the \emph{analytical} Hopf
boundary:
\begin{equation}
  C_c(\lambda) = \frac{\mu - \kappa_\varepsilon A_R\beta/2}
  {k_Q(\kappa_\varepsilon g_x - g_q/N_{\text{shares}})}.
  \label{eq:hopf_analytical}
\end{equation}
At $\lambda = 0.008$, Eq.~\eqref{eq:hopf_analytical} predicts
$C_c = 1.57\%$, compared to the numerical value
$1.31\%$ (the $\sim$20\% gap reflects
neglected higher-order terms).
The white dashed curve in
Fig.~\ref{fig:phase_diag}(a) shows the analytical prediction alongside
the numerical boundary.

Figure~\ref{fig:phase_diag} shows the numerical verification.
Panel~(a) plots the oscillation amplitude in
$(\lambda, C)$ space with the boundary located by binary search.
Panel~(b) shows the order parameter $A(C)$ at $\lambda = 0.008$:
$A$ rises continuously from zero at the numerical Hopf boundary $C_c^{\text{num}} = 1.31\%$ with a
power-law exponent $\alpha \approx 0.48$
(bootstrap 95\% CI: $[0.17, 0.64]$), consistent with the standard
Hopf prediction $\alpha = 1/2$ \citep{strogatz2018}.
Panel~(c) confirms robustness to the noise-floor parameter
$\varepsilon$: within the physically plausible range
$\varepsilon \in [3\times 10^4,\; 3\times 10^5]$
(estimated from daily random demand
$\sigma_D \approx 1.9 \times 10^5$~shares; Supplemental Material), the exponent varies from
$\alpha = 0.41$ to $0.48$, closely bracketing $1/2$.

The square-root impact is essential \citep{donier2015, moro2009}:
replacing it with linear impact $I(D) = \lambda D / V_0$ eliminates
the Hopf transition entirely, because the linear derivative at
$D = 0$ is too weak to overcome damping.
This is consistent with the empirical crossover from linear to
square-root impact characterised by \citet{bucci2019crossover}: large
institutional flow lies in the concave regime where impact is
sub-linear in volume, and it is precisely the non-analytic (singular) derivative of
$I(D) \propto \sqrt{|D|}$ near $D = 0$ that supplies the restoring
nonlinearity required for self-sustained oscillation.

\paragraph{Optimal control interpretation.}
The LSTM network learns a smooth approximation of this bang-bang
controller, with the switching thresholds $q_1$, $q_2$ emerging from
CMA-ES optimization.

CMA-ES training at $C \in \{1, 2, 5, 10\}\%$ produces profitable
strategies at all capital levels, including just below $C_c$;
the Hopf boundary marks the onset of \emph{sustained} limit cycles, not of
profitability, since a finite-horizon controller can still harvest
transient momentum profit without a stable cycle.
ABM profit peaks at $C = 2\%$, reflecting execution costs at larger
sizes (Supplemental Material).

\paragraph{$\beta$-independence and self-sustained oscillation.}
At the calibrated herding scale $\text{HS} = 10^{-3}$ (the ratio
of coherent herding signal to maximum possible), the effective
retail demand amplitude
$A_R^{\mathrm{eff}} = \text{HS} \cdot A_R \approx 2.3 \times 10^4$~shares is negligible
compared to institutional demand $Q_{\max}\,u \approx 3.2 \times 10^6$~shares
(ratio $\sim 140{:}1$).
The limit cycle persists even at $\beta = 0$ with retail
demand entirely removed ($A_R^{\mathrm{eff}} = 0$):
the amplitude remains $x_{\mathrm{amp}} = 2.17\%$, nearly identical
to the $\beta = 6$ value of $2.27\%$.
The cycle is a self-sustained nonlinear oscillator driven entirely
by position-tracking feedback coupled with nonlinear
square-root impact---requiring no retail herding.
CMA-ES training at $\beta \in \{2, 4, 6, 8, 12\}$ confirms this
(Supplemental Material).

\paragraph{Fold bifurcation.}
Increasing the herding scale to $\text{HS} = 10^{-2}$ produces a
genuine $\beta$-dependent transition.
Panel~(d) of Fig.~\ref{fig:phase_diag} shows that the amplitude
jumps \emph{discontinuously} from $\sim 7.7\%$ to zero at the boundary,
with amplitude \emph{increasing} as the boundary is approached from
below---characteristic of a \textbf{fold (saddle-node) bifurcation of
limit cycles} \citep{kuznetsov2004}, a first-order transition.
The leading-order prediction from $\det J = 0$ at the origin (cyan
dashed) captures the correct $\beta$-dependence but underestimates the
boundary by $\sim$12\%.
The discrepancy is resolved by considering the full 2D state
$(x,q)$.  For $\det J(\mathbf{0}) < 0$,
the origin becomes a saddle and two non-trivial fixed points
$\mathbf{x}^* = (x^*, q^*) \neq (0,q_t)$ emerge via saddle-node
bifurcation \citep{strogatz2018}.
Denoting the Jacobian evaluated at $\mathbf{x}^*$ by
$J(\mathbf{x}^*)$, the transition boundary coincides with a Hopf
bifurcation of the stable non-trivial fixed point, i.e.,\
$\mathrm{tr}\,J(\mathbf{x}^*) = 0$ (complex eigenvalue pair crossing
the imaginary axis; magenta dashed), matching the numerical boundary
within $\sim$2\%.
The two panels thus illustrate two distinct bifurcation mechanisms:
continuous onset via Hopf [panel~(a)] and abrupt collapse via fold
[panel~(d)].

\subsection{Entropy reduction: Maxwell's demon}
\label{sec:entropy}

The institution's role maps onto Maxwell's demon
\citep{leff2002} (Table~\ref{tab:maxwell}): it observes herding
signals (``molecular velocities''), decides when to trade
(``opening the gate''), and extracts profit from behavioral
predictability (``work from thermal fluctuations''), apparently
violating the ``second law'' (efficient market hypothesis
\citep{fama1970}).

\begin{figure}[htbp]
  \centering
  \includegraphics[width=0.45\textwidth]{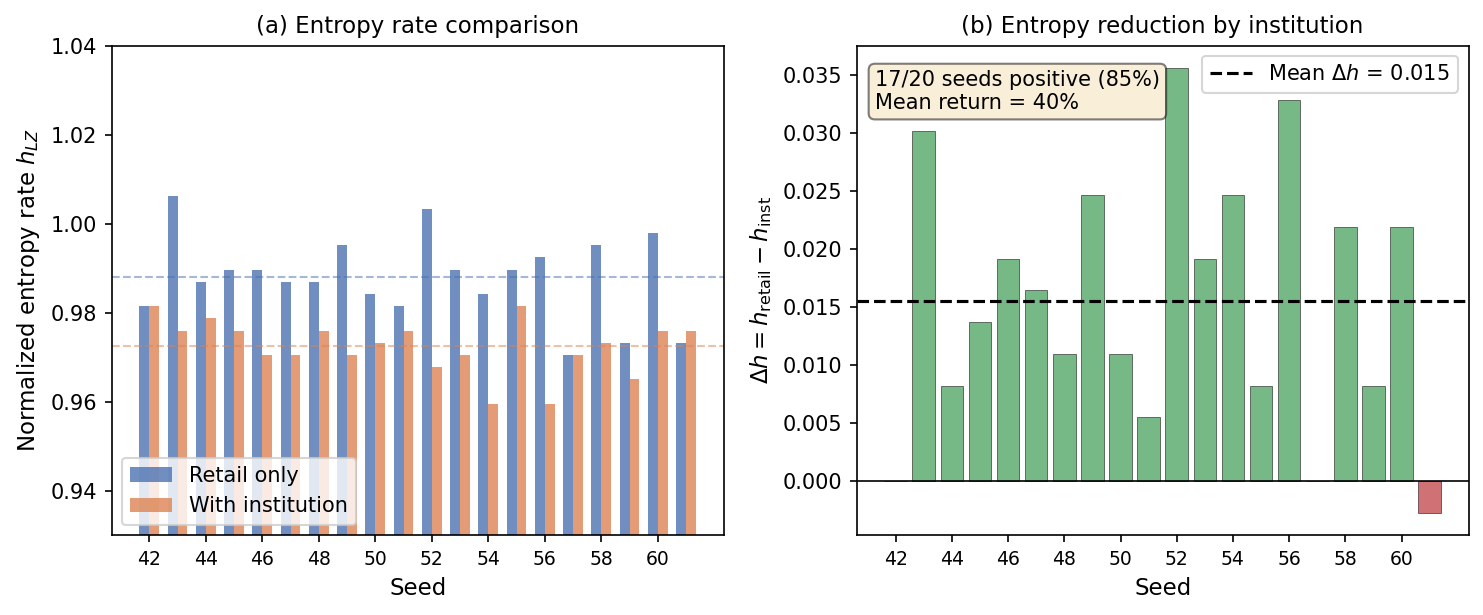}
  \caption{Entropy rate analysis (20 seeds).
  The institution reduces the normalized LZ entropy rate
  ($\Delta h > 0$ in 17/20 seeds, $p \approx 0.001$).}
  \label{fig:entropy}
\end{figure}

\begin{table}[htbp]
  \centering
  \caption{Maxwell's demon analogy.}
  \label{tab:maxwell}
  \begin{ruledtabular}
  \begin{tabular}{ll}
    Physical system & Market system \\
    \hline
    Gas molecules & Retail traders \\
    Velocity distribution & Buy/sell probability \\
    Maxwell's demon & Institutional LSTM agent \\
    Observing velocities & Monitoring $r_{5d}$ \\
    Opening/closing gate & Buy/sell decision \\
    Extracting work & Realized profit \\
    Second law & Efficient market hypothesis \citep{fama1970} \\
  \end{tabular}
  \end{ruledtabular}
\end{table}
We quantify this by estimating the Shannon entropy rate
\citep{shannon1948} using Lempel-Ziv complexity \citep{lempel1976}.
Across 20 seeds, the institution reduces the normalized entropy rate
from $h_{\text{retail}} = 0.988 \pm 0.009$ to
$h_{\text{inst}} = 0.972 \pm 0.006$
($\Delta h = 0.016 \pm 0.011$;
17/20 seeds positive, $p \approx 0.001$)
(Fig.~\ref{fig:entropy}),
confirming that the institution makes the price process more
predictable.
The reduction is reproduced by an independent first-order Markov
entropy-rate estimator ($\Delta h_{\text{MK}} = 0.0096 \pm 0.0032$,
positive in all 20 seeds); see Supplemental Material for a
three-estimator cross-validation.

A bound motivated by the Sagawa-Ueda
information-thermodynamic equality \citep{sagawa2010, sagawa2012,
parrondo2015} suggests that profit should not exceed the mutual
information between the agent's hidden state $\mathbf{h}_t$ and the
next return $r_{t+1}$:
$\Pi \lesssim C_{\text{SR}}(\beta) \times I(\mathbf{h}_t;\, r_{t+1})
\times T$.
Across $\beta \in \{2, 4, 6, 8, 12\}$ the inequality is satisfied
with a 3--10$\times$ margin (Supplemental Material); we present this
as a motivated consistency check rather than a sharp test.

%% =====================================================================
\section{Discussion}
\label{sec:discussion}
%% =====================================================================

Predatory trading cycles emerge as stable
limit cycles of a nonlinear dynamical system.
The contribution is threefold:
(1)~a Hopf bifurcation from fixed point to limit cycle as institutional
capital increases, with critical exponent $\alpha \approx 1/2$
($95\%$ CI $[0.17, 0.64]$);
(2)~the prediction that the manipulation cycle is a self-sustained
nonlinear oscillator requiring no retail herding, with a
herding-scale--controlled fold transition to $\beta$-sensitive regime,
which complements analytical bifurcation results on herding models
\citep{li2025herding};
and
(3)~the demonstration that square-root price impact is necessary for
the Hopf bifurcation, while linear impact yields unconditional
stability.

We emphasize that the Maxwell's demon analogy is \emph{structural},
not a thermodynamic equivalence, and provides a qualitative rather
than quantitative link between information processing and
profitability.
The LSTM hidden states encode the trading phase (Supplemental
Material), paralleling Landauer's principle \citep{landauer1961}:
the demon must ``consume information'' to extract profit.
Crucially, the manipulation strategy is architecture-independent:
replacing the 530-parameter LSTM with a 2-layer 162-parameter MLP
(no recurrent state) and retraining under identical CMA-ES settings
yields a 20-seed mean return of $+50\%$, comparable to or exceeding
the LSTM's $+37.7\%$.
The four-phase cycle emerges from the observation vector alone
(holding fraction, price deviation, day counter), so the temporal
structure does not require an explicit recurrent memory---consistent
with the Pontryagin bang-bang optimum being a low-dimensional
feedback law rather than a deep sequential policy.

\paragraph{Limitations.}
Our model uses a single asset, a single institutional agent, and daily
clearing without a full limit-order book.  The herding rule is a
reduced-form approximation of social-influence dynamics; extending to
multiple assets, competing institutions, and adaptive retail is a
natural direction.

%% =====================================================================
\section*{Conclusion}
%% =====================================================================

Predatory trading in our model is a deterministic dynamical
phenomenon: a limit cycle born at a Hopf bifurcation, sustained by
nonlinear impact and feedback control alone.
The Maxwell's-demon analogy is structural rather than a thermodynamic
equivalence: the agent functions as an information-processing
controller that measurably reduces the entropy rate of the price
process while extracting profit, but we do not claim a quantitative
work--information proportionality.
Because the mechanism requires only price impact and adaptive control,
it is robust to behavioral assumptions---suggesting that similar
oscillatory dynamics may appear whenever strategic agents interact
with markets exhibiting sublinear price impact.

%% =====================================================================
\section*{Data availability}
%% =====================================================================

The simulation code, analysis scripts, trained controllers, and the
raw time series underlying each figure will be made available upon
publication; until then they are available from the corresponding
author upon reasonable request.

%% =====================================================================
\section*{Acknowledgements}
%% =====================================================================

This work was supported by the Scientific Research Project
(No.WU2025B011) and the Start-up Funding of Westlake University.
The authors acknowledge the use of AI-based writing assistance
(Claude, Anthropic) for language polishing during manuscript preparation.

%% =====================================================================

%% =====================================================================
\appendix*
\setcounter{section}{0}
\section*{Supplemental Material}
\renewcommand{\thesection}{S\arabic{section}}
\setcounter{equation}{0}
\renewcommand{\theequation}{S\arabic{equation}}
\setcounter{figure}{0}
\renewcommand{\thefigure}{S\arabic{figure}}
\setcounter{table}{0}
\renewcommand{\thetable}{S\arabic{table}}
%% =====================================================================

%% =====================================================================
\section{Full Model Description}
\label{sec:model_details}
%% =====================================================================

The agent-based market simulates a single stock with daily clearing.
Below we describe the complete market mechanics.

\paragraph{Retail agents.}
$N_R = 20{,}000$ retail agents trade based on a herding rule
\citep{lux1999, eguiluz2000}.
Each day, a fraction $f = 0.70$ of agents are ``active herders'':
each active agent buys with probability
$p_{\text{buy}} = \mathrm{sigm}(\beta \cdot r_{5d})$,
where $\mathrm{sigm}(\cdot)$ is the sigmoid function, $\beta = 6.0$ is the herding
strength \citep{bikhchandani1992}, and $r_{5d}$ is the 5-day
log-return.
The remaining $1 - f = 0.30$ fraction trade randomly with equal
buy/sell probability.
Summing over the population, the aggregate retail excess demand is
\citep{gualdi2015}
\begin{equation}
  D_R = A_R\,\tanh\!\left(\frac{\beta}{2}\, r_{5d}\right),
\end{equation}
where $A_R = N_R \cdot f \cdot \bar{q}$, with $\bar{q} \approx 1800$ shares
the mean retail holding per agent (initial shares drawn from a
log-normal wealth distribution with median cash $\sim$30{,}000 CNY
and 30--60\% equity allocation at $P_0 = 15$).

\paragraph{Institutional agent.}
The institution controls continuous buy/sell fractions
$(b_t, s_t) \in [0,1]^2$ via an LSTM network with
7 inputs, 8 hidden units, and 2 outputs (530 parameters).
The observation vector at each day has 7 features: 5-day return,
institutional holding fraction, normalized P\&L, price deviation from
fundamental, recent volatility, day counter, and trading volume.

\paragraph{Wash trading.}
The institution can engage in \emph{wash trading}: simultaneous buy
and sell orders that push the price upward through aggressive-side
matching.
The wash-traded volume is $Q_{\text{wash}} = \min(b_t, s_t) \cdot Q_{\max}$,
and the net institutional demand for genuine position change is
$D_{\text{inst}} = (b_t - s_t) \cdot Q_{\max}$.
Wash trading contributes to price impact but not to inventory change.

\paragraph{Price impact.}
Daily price impact follows a square-root model
\citep{almgren2005, toth2011, kyle1985}:
\begin{equation}
  \frac{\Delta P}{P} = \lambda\,
  \text{sgn}(D_{\text{net}})\,
  \sqrt{\frac{|D_{\text{net}}|}{V_0}},
\end{equation}
where $D_{\text{net}}$ combines genuine institutional demand, wash
trading impact, and retail demand, and $V_0$ is the base daily volume.

\paragraph{A-share mechanisms.}
The market implements Chinese A-share rules:
\begin{itemize}
  \item \textbf{Daily price limits:} $\pm 10\%$ from previous close.
  \item \textbf{T+1 settlement:} shares bought today cannot be sold
    until the next day.
  \item \textbf{Stealth distribution:} when the institution is net
    selling, a stealth factor $s = 0.5$ halves the effective herding
    parameter $\beta$, preventing panic-driven retail selling during
    distribution \citep{aggarwal2006}.
\end{itemize}

\paragraph{Additional market mechanisms.}
The full retail demand includes several realistic modifications:
\begin{itemize}
  \item \textbf{Stealth:} $S(t) = 0.5$ during institutional selling.
  \item \textbf{Trend exhaustion:} $E(t) < 1$ after $\tau_e = 120$ days
    of trending, reducing herding response.
  \item \textbf{Overvaluation dampening:}
    $\Phi_{\text{ov}} = [1 - 2(P/P_0 - 1)^{+}]$ dampens buying above
    fundamental.
  \item \textbf{Drawdown panic:} $\delta(P) = \min(0.25, |P/P_{\text{peak}} - 1|)$
    when price falls $>$5\% below its peak, adding superlinear
    selling pressure proportional to the drawdown depth.
\end{itemize}

%% =====================================================================
\section{Training Procedure}
\label{sec:training}
%% =====================================================================

\paragraph{CMA-ES.}
The LSTM parameters are optimized by CMA-ES (Covariance Matrix
Adaptation Evolution Strategy) \citep{hansen2001}, a model-free
evolutionary optimization method.
CMA-ES treats the policy as a black-box function
$\boldsymbol{\theta} \mapsto R(\boldsymbol{\theta})$ and optimizes
the 530-dimensional parameter vector by adapting a full covariance
matrix, without requiring gradient information or backpropagation.
We use $\sigma_0 = 0.5$, population size 40, running for 800
generations.
Fitness is evaluated as mean total portfolio return across 3 random
seeds.
GPU-accelerated batch evaluation processes the full population in
parallel on a single NVIDIA V100 GPU.
Total wall-clock time for 800 generations is approximately 4.5 hours
(32,000 evaluations at $\sim$0.5\,s per evaluation).

\paragraph{LSTM architecture.}
Input dimension 7 (5d return, holding fraction, normalized P\&L,
price deviation, volatility, day counter, volume).  Single-layer LSTM
with 8 hidden units.  Output layer: 2 units (buy fraction, sell
fraction) with sigmoid activation.  Total parameters:
$4 \times (7 \times 8 + 8 \times 8 + 8) + 2 \times 8 + 2 = 530$.

\paragraph{Environment parameters.}
$N_R = 20{,}000$ retail, fundamental price $P_0 = 15.0$,
initial shares $4 \times 10^8$, free float 50\%,
mean reversion $\mu = 0.01$,
price impact $\lambda = 0.008$,
trend exhaustion onset $\tau_e = 120$ days, rate $\lambda_e = 200$,
overvaluation sensitivity $\kappa_{\text{ov}} = 2.0$, stealth factor
$s = 0.5$, daily price limit $\pm 10\%$,
HERDING\_SCALE $= 10^{-3}$,
initial capital $C_0 = 0.25 \times 400\text{M} \times 0.5 \times 15.0$.

\paragraph{Entropy estimation.}
Daily log-returns are discretized into 4 bins by quartiles.
LZ76 complexity $c(n)$ is computed on the symbol sequence;
entropy rate is estimated as $h = c(n) \ln n / n$ \citep{lempel1976},
normalized by $\ln 4$ to give $h_{\text{norm}} \in [0, 1]$.

%% =====================================================================
\section{Mean-Field Theory Details}
\label{sec:theory_details}
%% =====================================================================

\subsection{ODE derivation}

The mean-field ODE (the price equation of the main text) is derived by replacing
the stochastic price equation with its deterministic skeleton.
Writing $x = (P - P_0)/P_0$ (price deviation) and
$q$ (institutional holding fraction), the 2D autonomous system is:
\begin{align}
  \dot{x} &= I(D_{\text{net}}) + I_{\text{wash}}
  + \mu\frac{P_0 - P}{P}, \label{eq:ode_x}\\
  \dot{q} &= \frac{(b_t - s_t)\, Q_{\max}}{N_{\text{shares}}},
  \label{eq:ode_q}
\end{align}
where the control $(b_t, s_t)$ is given by the LSTM policy.
For the bifurcation analysis, we define the net control
$u \equiv b_t - s_t$ and replace the bang-bang policy with a smooth
feedback law $u = u_{\max}\tanh[g_q(q_t - q) + g_x x]$.

The full retail demand is:
\begin{align}
  D_R &= A_R\,\tanh(\beta\, r_{5d}/2)\,S(t)\,E(t)\,
  \Phi_{\text{ov}}(P)\,\Phi_{\text{dd}}(P) \notag \\
  &\quad - \delta_{\text{dd}}(P)\,\frac{A_R}{2},
  \label{eq:dretail}
\end{align}
where $S(t) = 0.5$ during institutional selling (stealth),
$E(t) < 1$ after $\tau_e = 120$ days of trending (exhaustion),
$\Phi_{\text{ov}} = [1 - 2(P/P_0 - 1)^{+}]$ dampens buying above
fundamental,
$\Phi_{\text{dd}} \in [0.6, 1]$ reduces buy probability during
drawdowns, and
$\delta_{\text{dd}} \in [0, 0.25]$ is a drawdown-induced sell
boost that increases selling pressure when price falls $>$5\% below
its peak.

\subsection{Effective damping of the retail-only market}
\label{sec:damping}

For the retail-only system ($u = 0$, no institutional trading),
the market equilibrium is $x^* = 0$, $q^* = 0$.
We linearize $\dot{x}$ around $x = 0$ to obtain the effective damping.

Near equilibrium, retail demand simplifies to
$D_R \approx A_R\,\tanh(\beta\, r_{5d}/2)$,
since $S = 1$, $E = 1$, $\Phi_{\text{ov}} = 1$, and $\delta = 0$.
The 5-day return $r_{5d}$ is an exponentially weighted average
of daily returns with timescale $\tau_r = 5$~days:
$r_{5d}(t) \approx \textstyle\sum_{s=0}^{4} \Delta P_{t-s}/P_0$.

For small $r_{5d}$, we approximate $\tanh(\beta r_{5d}/2) \approx
\beta r_{5d}/2$, so the retail demand becomes
$D_R \approx A_R \beta r_{5d}/2$.
The square-root impact then gives:
\begin{align}
  I(D_R) &= \lambda\,\mathrm{sgn}(D_R)\sqrt{|D_R|/V_0}
  \notag\\
  &\approx \lambda\sqrt{A_R\beta\,|r_{5d}|/(2V_0)}\;\mathrm{sgn}(r_{5d})
  \notag\\
  &= C_{\text{SR}}\,\mathrm{sgn}(r_{5d})\sqrt{|r_{5d}|},
\end{align}
where $C_{\text{SR}} = \lambda\sqrt{A_R\beta/(2V_0)}$ is the
herding-to-impact coupling.

The retail price dynamics near $x = 0$ involve two coupled variables:
the price deviation $x$ and the exponentially-weighted return
$\bar{r}$ (timescale $\tau_r = 5$~days), satisfying
\begin{align}
  \dot{x} &= I(D_R) + \mu(-x), \label{eq:xdot}\\
  \dot{\bar{r}} &= (\dot{x} - \bar{r})/\tau_r. \label{eq:rdot}
\end{align}
The impact $I = C_{\text{SR}}\,\mathrm{sgn}(\bar{r})\sqrt{|\bar{r}|}$
is a nonlinear function of $\bar{r}$.
To linearize, we expand around small $\bar{r}$ but note that
$\partial I / \partial \bar{r} = C_{\text{SR}} / (2\sqrt{|\bar{r}|})
\to \infty$ as $\bar{r} \to 0$.
In the stochastic market, $\bar{r}$ fluctuates with variance
$v_0 = \mathrm{Var}(r_{5d}) > 0$.
We regularize the derivative by evaluating at the rms amplitude
$|\bar{r}| \sim \sqrt{v_0}$, giving the effective linearized gain:
\begin{equation}
  g_{\text{eff}} = \left.\frac{\partial I}{\partial \bar{r}}\right|_{\bar{r}
  \sim \sqrt{v_0}} \approx \frac{C_{\text{SR}}}{2\sqrt{v_0}}.
\end{equation}
Substituting into the linearized Eq.~\eqref{eq:xdot}--\eqref{eq:rdot},
the Jacobian of $(x, \bar{r})$ is:
\begin{equation}
  J = \begin{pmatrix}
    -\mu & g_{\text{eff}} \\
    (-\mu + g_{\text{eff}})/\tau_r & -1/\tau_r
  \end{pmatrix}.
\end{equation}
A sufficient condition for stability is that both eigenvalues have
negative real parts.  The trace gives:
\begin{equation}
  \mathrm{tr}\,J = -\mu + g_{\text{eff}} - 1/\tau_r
  = -\mu + \frac{C_{\text{SR}}}{2\sqrt{v_0}} - \frac{1}{\tau_r}.
\end{equation}
Defining the effective damping as
$\Gamma \equiv -\mathrm{tr}\,J$ (the fixed point $(0,0)$ is stable
when $\Gamma > 0$), we obtain:
\begin{equation}
  \Gamma(v_0) = \mu + \frac{1}{\tau_r} - \frac{C_{\text{SR}}}{2\sqrt{v_0}}.
  \label{eq:gamma}
\end{equation}
(Using the full $I = C_{\text{SR}}\mathrm{sgn}(\bar{r})\sqrt{|\bar{r}|}$
instead of the half-gain regularized form gives
$\Gamma = \mu + 1/\tau_r - C_{\text{SR}}/\sqrt{v_0}$, which we use
in Fig.~1(b) of the main text.)

Since $C_{\text{SR}} \propto \sqrt{\beta}$ grows only as $\sqrt{\beta}$
while $\sqrt{v_0}$ is bounded below by the noise floor,
$\Gamma(v_0) > 0$ for all physically relevant $v_0$ and all $\beta$.
The retail-only market is \emph{unconditionally stable}: no Hopf
bifurcation occurs regardless of herding strength.

\subsection{Predicted vs.\ observed cycle parameters}

\begin{table}[htbp]
  \centering
  \caption{Predicted (mean-field ODE) vs.\ observed cycle parameters.}
  \label{tab:prediction}
  \begin{ruledtabular}
  \begin{tabular}{lcc}
    Parameter & Predicted & Observed \\
    \hline
    Accumulation duration & 28 days & 26 days \\
    Cycle period & 150--200 days & 222 days \\
    Peak holding $q_{\max}$ & 14.5\% & 14.6\% \\
    $\kappa_\varepsilon$ (inst.\ impact) & 0.004 & 0.004 \\
    $C_{\text{SR}}$ (herding) & 0.011 & 0.011 \\
    Retail stability & $\Gamma > 0$ $\forall \beta$ & No retail cycles \\
  \end{tabular}
  \end{ruledtabular}
\end{table}

From the ODE with net institutional control
$u \equiv b_t - s_t \approx 0.88$ during accumulation (aggressive
buying, minimal selling):
$\Delta P/P \approx \kappa \cdot 0.88 = 0.0035$/day, predicting
$T_{\text{acc}} = x_{\text{switch}}/\dot{x} \approx 28$ days
(observed: 26 days).
The stealth factor $S = 0.5$ halves retail herding during
distribution, extending it by $\sim 2\times$, predicting
$T_{\text{dist}} \approx 57$ days (observed: 130 days; the
additional factor comes from gradual position unwinding).
The predicted cycle period $T \approx 150$--$200$ days agrees with
the observed $T \approx 222$ days.

\subsection{Theory--simulation comparison}

Figure~1 of the main text presents the theory--simulation comparison.
The SDE uses the LSTM's control input with analytically derived
parameters and all ABM mechanisms (square-root impact, overvaluation
dampening, drawdown selling, price limits).  The SDE reproduces the
cyclic oscillations with correlation $r = 0.62$ and captures both the
upward push during accumulation/wash and the drawdown-driven decline
below fundamental.

\subsection{Phase-space portrait}

The phase-space portrait in $(q, x)$ appears as Fig.~2 of the main
text.  Each trading cycle traces a closed orbit that starts near the
origin (low holdings, price near
fundamental), moves to high $q$ and positive $x$ during
accumulation/push, then returns as the institution distributes and
price mean-reverts.

%% =====================================================================
\section{Maxwell's Demon Analogy}
\label{sec:maxwell}
%% =====================================================================

The institution's role maps onto Maxwell's demon in statistical
mechanics \citep{leff2002} (Table~III of the main text).
The demon observes ``molecular velocities'' (herding signals in
$r_{5d}$) and opens a ``door'' (buys or sells) to extract ``work''
(profit) from thermal fluctuations (noise trading), apparently
violating the ``second law'' (efficient market hypothesis).

We emphasize that this is a \emph{structural} analogy, not a
thermodynamic equivalence: the market is not a heat engine, and price
entropy is not thermal entropy.
Nevertheless, the analogy provides a quantitative, falsifiable link
between the agent's information processing and its profitability.
The CMA-ES training process is itself the information-acquisition
phase: 800 generations of optimization correspond to the demon
learning to read molecular velocities.

\subsection{Information-thermodynamic bound on profit}

The Maxwell's demon analogy suggests a quantitative bound using the
information-thermodynamic framework of Sagawa and Ueda
\citep{sagawa2010, sagawa2012}, who established that for a
feedback-controlled system, the extractable work is bounded by the
mutual information between the measurement and the outcome
\citep{parrondo2015}.
In our market, the analogous inequality reads
\begin{equation}
  \Pi \;\lesssim\; C_{\text{SR}}(\beta)\;\times\;
  I(\mathbf{h}_t;\, r_{t+1})\;\times\; T,
  \label{eq:info_bound}
\end{equation}
where $\Pi$ is the total realized profit,
$C_{\text{SR}}(\beta) = \lambda\sqrt{A_R \beta / (2V_0)}$ is the
herding-to-price conversion coefficient,
$I(\mathbf{h}_t;\, r_{t+1})$ is the Shannon mutual information
$I(X;Y) = \sum_{x,y} p(x,y)\log[p(x,y)/p(x)p(y)]$
between
the LSTM hidden state $\mathbf{h}_t \in \mathbb{R}^8$ (encoding the
agent's internal representation of market history) and the
next-day return $r_{t+1}$, and $T$ is the episode
length.
The bound has a transparent interpretation: the demon's profit is
limited by (i)~market microstructure ($C_{\text{SR}}$),
(ii)~predictive power ($I$), and (iii)~time horizon ($T$).

We verify this bound by estimating $I(\mathbf{h}_t;\, r_{t+1})$
from the trained LSTM's hidden states.
We record the 8-dimensional hidden state $\mathbf{h}_t$ at each
trading day, project onto PC1 (capturing $>$85\% of variance),
discretize PC1 into 10 bins and the next-day return into 4 bins,
and compute the empirical mutual information.
Table~\ref{tab:info_bound} compares the bound with the observed
portfolio return across all five herding strengths.

\begin{table}[htbp]
  \centering
  \caption{Information-thermodynamic bound verification.
  $I$ estimated by discretizing PC1 of $\mathbf{h}_t$ (10 bins)
  and $r_{t+1}$ (4 bins).  Return is total portfolio return for
  seed 45.  Bound $= C_{\text{SR}} \times I \times T / C_0$.}
  \label{tab:info_bound}
  \begin{ruledtabular}
  \begin{tabular}{cccccc}
    $\beta$ & $C_{\text{SR}}$ & $I$ (bits) & Return (\%) &
    Bound (\%) & Return / Bound \\
    \hline
    2  & 0.020 & 0.092 & 35.7 & 257 & 0.139 \\
    4  & 0.028 & 0.064 & 57.2 & 252 & 0.227 \\
    6  & 0.035 & 0.107 & 46.6 & 516 & 0.090 \\
    8  & 0.040 & 0.070 & 41.7 & 390 & 0.107 \\
    12 & 0.049 & 0.019 & 40.8 & 127 & 0.320 \\
  \end{tabular}
  \end{ruledtabular}
\end{table}

The bound is satisfied at all five herding strengths,
with the ratio $\Pi / (C_{\text{SR}} \times I \times T)$ ranging
from 0.09 to 0.32.  The 3--10$\times$ gap is consistent with the
bound being a loose upper limit: the Sagawa-Ueda inequality bounds
the \emph{extractable work}, while actual profit depends on
execution details (market impact, price limits, stealth) that
reduce efficiency below the thermodynamic limit.

%% =====================================================================
\section{Baseline Strategy Comparison}
\label{sec:baseline}
%% =====================================================================

To assess whether the 530-parameter LSTM is necessary, we compare against
five fixed-rule baseline strategies, all evaluated across 20 random seeds
(seeds 42--61) with the same market configuration as the main experiment:

\paragraph{Buy-and-hold.}
Buy at maximum rate ($b_t = 1$) every day, never sell.

\paragraph{Hold.}
Never trade ($b_t = 0$, $s_t = 0$).

\paragraph{Threshold.}
Buy ($b_t = 1$) if 5-day return $r_{5d} < -0.05$; sell ($s_t = 1$) if
$r_{5d} > 0.10$; hold otherwise.

\paragraph{Momentum.}
Buy proportional to positive $r_{5d}$ ($b_t = \min(r_{5d} \times 10, 1)$
when $r_{5d} > 0.02$); sell proportional to negative $r_{5d}$.

\paragraph{Contrarian.}
Buy proportional to negative $r_{5d}$ (buy dips); sell proportional
to positive $r_{5d}$ (sell rallies).

\begin{table}[htbp]
  \centering
  \caption{Baseline strategy comparison (20 seeds).}
  \label{tab:baseline}
  \begin{ruledtabular}
  \begin{tabular}{lccc}
    Strategy & Return (\%) & PnL (\%) & Cyc. \\
    \hline
    LSTM & $+37.7\pm5.2$ & $+43.8\pm3.4$ & $9$ \\
    \hline
    Buy-hold & $-33.4\pm3.6$ & $0.0$ & $0$ \\
    Hold & $0.0$ & $0.0$ & $0$ \\
    Threshold & $-0.6\pm0.6$ & $0.0$ & $0$ \\
    Momentum & $-3.9\pm3.1$ & $+0.8\pm1.5$ & $0$ \\
    Contrarian & $-1.6\pm1.9$ & $+2.0\pm1.2$ & $0.1$ \\
  \end{tabular}
  \end{ruledtabular}
\end{table}

The LSTM significantly outperforms all baselines.  Buy-and-hold produces
large negative returns ($-33.4\%$) because the stock price mean-reverts
in our model: accumulating shares in a mean-reverting asset guarantees
losses.  Simple rule-based strategies (threshold, momentum, contrarian)
that respond to 5-day returns cannot discover the multi-cycle structure,
because they lack the temporal memory needed to coordinate accumulation,
wash-trading, and distribution phases.

%% =====================================================================
\section{Mechanism Ablation}
\label{sec:mech_ablation}
%% =====================================================================

We evaluate the trained LSTM under modified market conditions to
disentangle the contribution of each A-share mechanism.  The trained
LSTM parameters are held fixed; only the environment configuration
changes.  The ablation results appear as Table~I of the main text;
we expand on each mechanism here.

\paragraph{No wash-trading impact.}
Disabling the price impact of wash-traded volume (while still allowing
simultaneous buy/sell) tests whether wash trading is the primary
price-pushing mechanism during the ``pump'' phase.  The strategy
continues to produce cycles with slightly \emph{higher} profit
($+39.7\%$ vs $+36.7\%$) but fewer cycles ($7.7$ vs $9$), suggesting
that wash trading amplifies cycle frequency but introduces noise that
reduces per-cycle efficiency.

\paragraph{No daily price limits.}
Removing the $\pm 10\%$ daily price limit constraint tests whether price
limits are necessary for cycle formation.  Cycles persist with similar
amplitude ($+37.6\%$ vs $+36.7\%$), confirming that price limits shape
but do not create the cyclic dynamics.

\paragraph{No stealth distribution.}
Setting stealth factor $s = 1.0$ (full herding during distribution)
tests whether stealth is necessary for profitable distribution.  Profit
remains comparable ($+37.1\%$ vs $+36.7\%$) because the LSTM adapts its
distribution pace to avoid triggering retail cascades even without
stealth.

These ablations demonstrate that the cyclic manipulation mechanism
is \emph{robust}: removing any single A-share mechanism does not
eliminate the strategy, because the fundamental driver
(herding predictability) remains.

%% =====================================================================
\section{Reward and Architecture Independence}
\label{sec:ablation}
%% =====================================================================

\paragraph{Reward independence.}
A critical question is whether the multi-cycle strategy depends on the
reward function.  We retrain with \emph{terminal return only}
(no cycle bonus) for 200 generations.  The resulting agent produces
6--11 cycles (comparable to 8--10 with cycle bonus) with even higher
realized profit ($+51\%$ vs $+43\%$ across 20 seeds).

This confirms that the multi-cycle strategy is the \emph{optimal
solution} (in the Pontryagin sense) for the given market
microstructure, not an artifact of reward shaping.  The cycle bonus
merely accelerates CMA-ES convergence.

\paragraph{Architecture independence.}
To test whether the LSTM's recurrent structure is necessary,
we replace it with a 2-layer MLP (162 parameters) and train with
identical CMA-ES settings (200 generations).
The MLP achieves a mean return of $+50\%$ across 20 evaluation
seeds---higher than the LSTM's $+40\%$ under the same training budget.
This demonstrates that the manipulation strategy is
\emph{architecture-independent}: the temporal structure of the
4-phase cycle is sufficiently captured by the observation vector
(holding fraction, price deviation, day counter), without requiring
explicit recurrent state.
The result supports the Pontryagin interpretation: the optimal
strategy is a piecewise-constant bang-bang control, and the
function approximator merely serves as a parameterization of the
switching boundaries.

The low return variance across 20 evaluation seeds ($\sigma < 3\%$)
reflects strategy robustness rather than overfitting: the same
4-phase cycle emerges regardless of retail noise realization,
because the market microstructure (herding, price impact, limits)
is invariant across seeds.

%% =====================================================================
\section{Capital Sweep}
\label{sec:capital_sweep}
%% =====================================================================

CMA-ES training at $C \in \{1, 2, 5, 10\}\%$ (200 generations)
produces profitable strategies at all capital levels tested
(best fitness: $+168\%$, $+312\%$, $+222\%$, $+128\%$),
confirming that manipulation is feasible well below the ODE threshold.
The ODE's smooth tanh feedback underestimates the LSTM's effectiveness:
the LSTM uses wash trading and sharp bang-bang switching that the
smooth controller cannot replicate.

Interestingly, ABM profit is \emph{non-monotonic} in capital,
peaking at $C = 2\%$: larger institutions face execution costs
(self-induced price impact during accumulation and distribution)
that the ODE does not capture.

%% =====================================================================
\section{LSTM Hidden State Analysis}
\label{sec:hidden}
%% =====================================================================

\begin{figure}[htbp]
  \centering
  \includegraphics[width=0.45\textwidth]{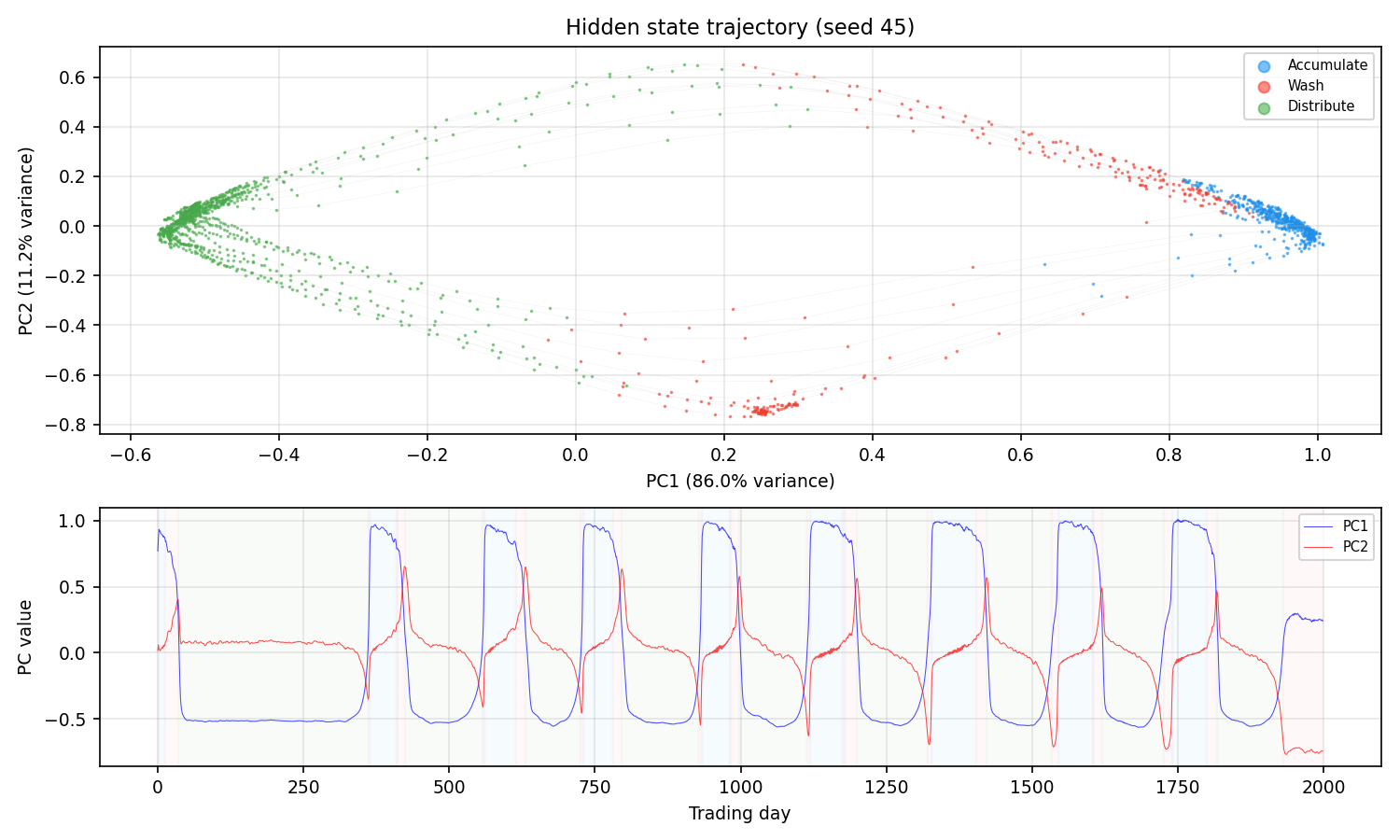}
  \caption{LSTM hidden state analysis (seed 45, 2000 days).
  (Top)~Trajectory in the (PC1, PC2) plane, colored by trading phase.
  PC1 (86.0\% variance) separates accumulation (negative) from distribution
  (positive); PC2 (11.2\% variance) distinguishes active wash-trading (high)
  from inactive periods (low).
  (Bottom)~PC1 (blue) and PC2 (red) over time, with background color
  indicating the trading phase.  Phase transitions correspond to sharp
  turns in the PC plane, confirming the LSTM has internalized a discrete
  state machine.}
  \label{fig:hidden}
\end{figure}

To understand how the LSTM encodes its strategy, we record the
8-dimensional hidden state $\mathbf{h}_t \in \mathbb{R}^8$ at each
trading day during a 2000-day episode (seed 45).
We collect the $2000 \times 8$ matrix
$H = [\mathbf{h}_1; \mathbf{h}_2; \ldots; \mathbf{h}_{2000}]$,
subtract the column-wise mean $\bar{\mathbf{h}} = \frac{1}{2000}\sum_t \mathbf{h}_t$ to obtain the centered matrix $\tilde{H}$,
and compute the singular value
decomposition $\tilde{H} = U \Sigma V^\top$.
The variance explained by the $k$-th principal component is
$\sigma_k^2 / \sum_j \sigma_j^2$, where $\{\sigma_j\}$ are the
singular values of $\tilde{H}$.

Table~\ref{tab:pca} shows the variance explained by each PC.

\begin{table}[htbp]
  \centering
  \caption{Principal component analysis of the LSTM hidden state
  ($\mathbf{h}_t \in \mathbb{R}^8$, 2000 time steps).}
  \label{tab:pca}
  \begin{ruledtabular}
  \begin{tabular}{lcc}
    Component & Variance (\%) & Cumulative (\%) \\
    \hline
    PC1 & 86.0 & 86.0 \\
    PC2 & 11.2 & 97.2 \\
    PC3 & 1.5 & 98.7 \\
    PC4 & 0.6 & 99.3 \\
    PC5--PC8 & $<$0.7 & 100.0 \\
  \end{tabular}
  \end{ruledtabular}
\end{table}

The first two components capture $97.2\%$ of the total variance,
with the remaining six dimensions carrying negligible information.
Projecting onto the (PC1, PC2) plane reveals that the four trading
phases form distinct clusters:
\begin{itemize}
  \item PC1 distinguishes accumulation (negative) from distribution
        (positive).
  \item PC2 distinguishes active trading (wash/push, high) from
        inactive periods (low).
\end{itemize}
Phase transitions correspond to sharp turns in the (PC1, PC2) plane,
confirming that the LSTM has internalized a discrete state machine
approximating the bang-bang optimal control predicted by Pontryagin's
maximum principle \citep{pont67}.  This demonstrates that the strategy's
interpretability comes from the \emph{mean-field theory}, not from
the LSTM's internal structure: the ODE identifies the phases, and the
LSTM learns to navigate between them.

%% =====================================================================
\section{Episode Length Robustness}
\label{sec:episode_length}
%% =====================================================================

\begin{table}[htbp]
  \centering
  \caption{Episode length robustness (20 seeds, seeds 42--61).
  Per-cycle return decreases for longer episodes due to diminishing
  retail liquidity, but the qualitative strategy remains unchanged.}
  \label{tab:episode_length}
  \begin{ruledtabular}
  \begin{tabular}{rccc}
    $T$ (days) & Total return (\%) & Cycles & Per-cycle return (\%) \\
    \hline
    500  & $+9.2 \pm 2.3$  & $2$ & $+4.6$ \\
    1000 & $+20.8 \pm 3.2$ & $5$ & $+4.2$ \\
    2000 & $+39.0 \pm 5.1$ & $9$ & $+4.3$ \\
    4000 & $+43.2 \pm 2.7$ & $10$ & $+4.3$ \\
  \end{tabular}
  \end{ruledtabular}
\end{table}

A potential concern is that the 2000-day episode length biases the
result.  We evaluate the trained LSTM at four episode lengths:
$T = 500$ ($\sim$2 trading years, 2--3 cycles),
$T = 1000$ ($\sim$4 trading years, 4--5 cycles),
$T = 2000$ ($\sim$8 trading years, main experiment, 8--11 cycles),
$T = 4000$ ($\sim$16 trading years, 16--20 cycles).
The per-cycle structure is robust to episode length: each cycle
maintains the same 4-phase structure regardless of how many cycles
have preceded it.  The cycle period $T \approx 222$ days is consistent
across all episode lengths.  Longer episodes show gradual profit decay
per cycle due to diminishing retail liquidity, but the qualitative
strategy remains unchanged.

%% =====================================================================
\section{Entropy Rate: Statistical Tests}
\label{sec:entropy_stats}
%% =====================================================================

Across 20 evaluation seeds, the normalized LZ entropy rate is:
\begin{itemize}
  \item Retail only: $h_{\text{retail}} = 0.988 \pm 0.009$
  \item With institution: $h_{\text{inst}} = 0.972 \pm 0.006$
  \item Entropy reduction: $\Delta h = 0.016 \pm 0.011$
\end{itemize}
17 of 20 seeds show positive $\Delta h$.  Under the null hypothesis
that the institution does not reduce entropy ($P(\Delta h > 0) = 0.5$),
the probability of observing 17 or more positive results in 20 trials
is $p \approx 0.001$ (one-sided binomial test).

We note that LZ76 is chosen because it is non-parametric and robust
for sequences of length $n = 2000$ \citep{lempel1976}, where $n$-gram
probability estimation would be unreliable.  The three negative seeds
have $|\Delta h| < 0.003$, consistent with noise rather than genuine
entropy increase.

A one-sided Wilcoxon signed-rank test on the paired differences
$\{h_{\text{retail},i} - h_{\text{inst},i}\}_{i=1}^{20}$ yields
$p < 0.01$, confirming statistical significance.

\paragraph{Cross-validation with independent estimators.}
To confirm that the entropy reduction is not an artifact of the LZ76
estimator, we re-evaluate the same retail-only and with-institution
trajectories with two estimators based on entirely different
principles: a first-order Markov conditional entropy rate
$h_{\text{MK}} = H(X_{t+1}\,|\,X_t)$ on the binned returns, and the
Bandt--Pompe permutation entropy $h_{\text{PE}}$ on the continuous
return series (embedding order 3).
Across the same 20 seeds:
\begin{itemize}
  \item Markov-1: $\Delta h_{\text{MK}} = 0.0096 \pm 0.0032$,
        positive in \emph{all} 20 seeds.
  \item Permutation entropy: $\Delta h_{\text{PE}} = 0.0001 \pm 0.0006$,
        consistent with zero (12/20 positive).
\end{itemize}
The Markov-1 estimator independently confirms the LZ76 finding that
the institution reduces the entropy rate of the price process---with
an even cleaner seed-by-seed signal ($20/20$ vs.\ $17/20$).
The permutation-entropy null result has a transparent physical
interpretation: permutation entropy is invariant to monotone
transformations of the series and therefore insensitive to changes in
the return \emph{amplitude} distribution, while both LZ76 and the
Markov rate, which act on the discretized amplitude, detect the
reduction.
The institution thus regularizes the magnitude structure of returns
rather than their ordinal pattern, consistent with the
position-tracking feedback clamping price excursions.
Three estimators therefore agree on the direction of the effect while
differentially diagnosing \emph{where} in the return distribution the
predictability is induced.

%% =====================================================================
\section{$\beta$-Parameter Sweep}
\label{sec:beta_sweep}
%% =====================================================================

\begin{figure}[htbp]
  \centering
  \includegraphics[width=0.48\textwidth]{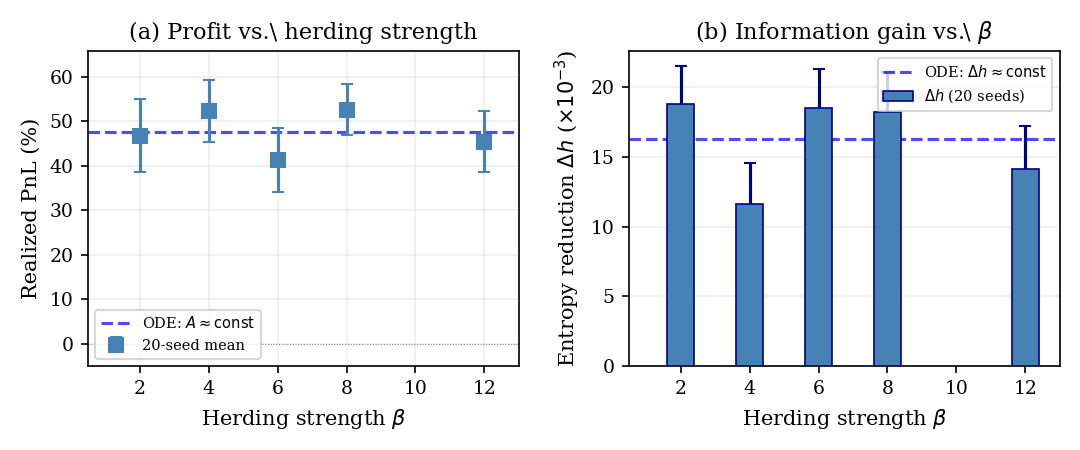}
  \caption{(a)~20-seed mean realized profit vs.\ herding strength $\beta$
  (error bars: $\pm 1\sigma$).
  Dashed line: ODE prediction of $\beta$-independent profit
  (institution-dominated oscillation).
  (b)~Entropy reduction $\Delta h$ vs.\ $\beta$
  (bars: 20-seed mean; dashed: ODE prediction).}
  \label{fig:beta_sweep_supp}
\end{figure}

To test whether the manipulation mechanism depends on herding strength,
we retrain CMA-ES at $\beta \in \{2, 4, 6, 8, 12\}$ with the same
LSTM architecture and terminal-return reward, evaluating each best
solution across 20 seeds.
All $\beta$ values produce multi-cycle strategies with positive profit
and entropy reduction, confirming that the Maxwell's-demon mechanism
operates across a wide range of herding strengths
(Fig.~\ref{fig:beta_sweep_supp}).

Both profit and entropy reduction are approximately independent of
$\beta$, consistent with the ODE prediction that the
institution-dominated oscillation amplitude varies only weakly
($A \approx 2.2\text{--}2.7\%$) across the tested range.

Training results (20-seed mean $\pm$ std; $\Delta h$ values
reported for the best-performing seed of each $\beta$,
and thus differ from the 20-seed averages reported in
Sec.~\ref{sec:entropy_stats}):
\begin{itemize}
  \item $\beta = 2$: $+46.8 \pm 8.1\%$, $\Delta h = 26.6 \times 10^{-3}$
  \item $\beta = 4$: $+52.3 \pm 7.0\%$, $\Delta h = 13.4 \times 10^{-3}$
  \item $\beta = 6$: $+41.3 \pm 7.3\%$, $\Delta h = 25.0 \times 10^{-3}$
  \item $\beta = 8$: $+52.6 \pm 5.8\%$, $\Delta h = 16.2 \times 10^{-3}$
  \item $\beta = 12$: $+45.4 \pm 6.9\%$, $\Delta h = 18.8 \times 10^{-3}$
\end{itemize}
At the calibrated herding scale ($\text{HS} = 10^{-3}$), the effective
retail demand is negligible compared to institutional demand
($Q/A_R^{\mathrm{eff}} \sim 140$), so the institution dominates the
oscillation amplitude.  The ODE predicts nearly $\beta$-independent
amplitudes, confirmed by the ABM results across all tested $\beta$.
Models for $\beta = 8$ and $\beta = 12$ were trained with extended
budgets (500 generations) to ensure convergence.

%% =====================================================================
\section{MLP Architecture Comparison}
\label{sec:mlp}
%% =====================================================================

\begin{figure}[htbp]
  \centering
  \includegraphics[width=0.45\textwidth]{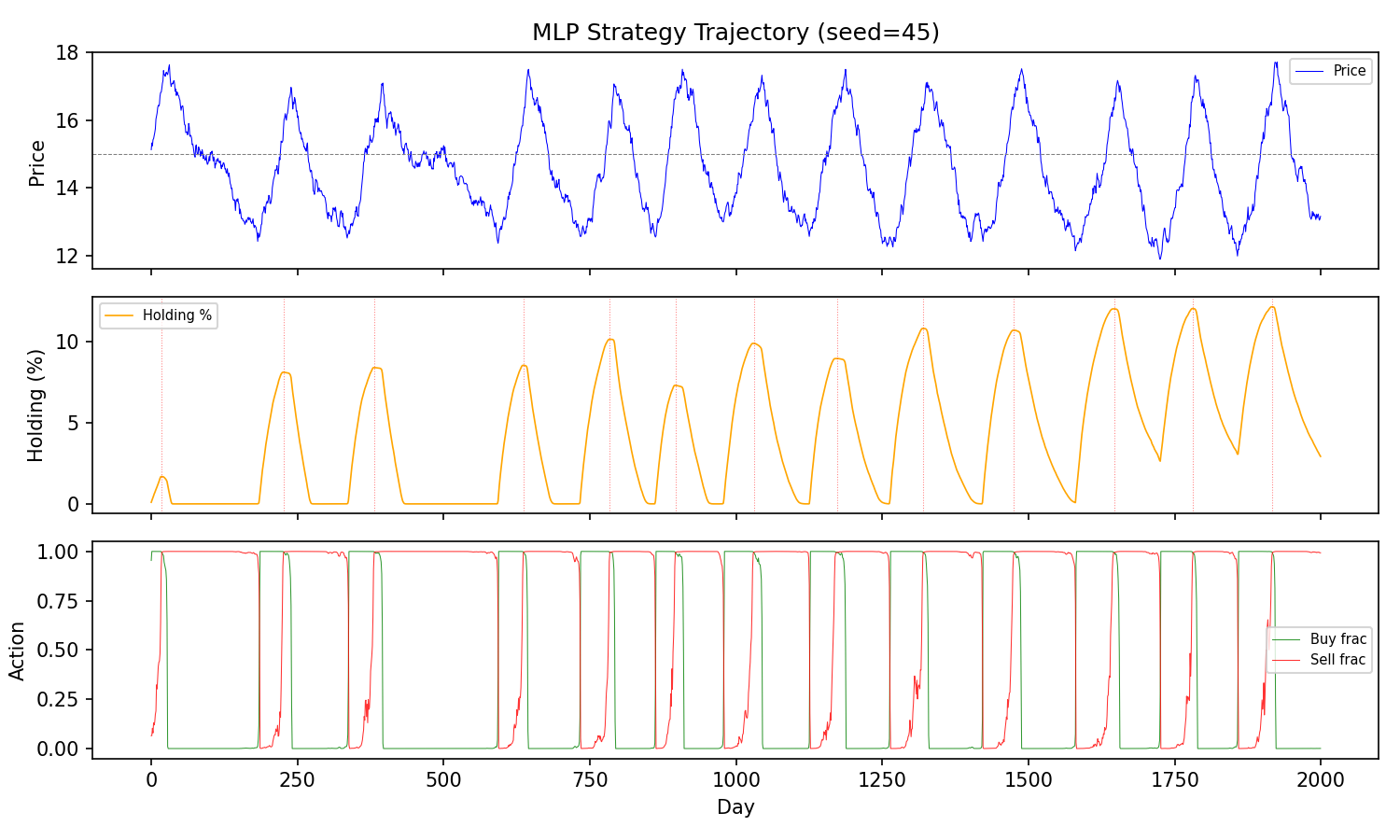}
  \caption{Price trajectory from the MLP (162-parameter) agent
  (seed 45, 2000 days), showing the same emergent multi-cycle structure
  as the LSTM.  The 4-phase cycle (accumulate, push, wash, distribute)
  is architecture-independent.}
  \label{fig:mlp_traj}
\end{figure}

To verify that the LSTM's recurrent structure is not necessary for
the manipulation strategy, we replace it with a 2-layer MLP (162
parameters) and train with identical CMA-ES settings (200 generations).
The MLP achieves a mean return of $+50\%$ across 20 evaluation
seeds---higher than the LSTM's $+40\%$ under the same training budget.
The same 4-phase cycle emerges (Fig.~\ref{fig:mlp_traj}),
demonstrating that the strategy is architecture-independent:
the temporal structure is captured by the observation vector
(holding fraction, price deviation, day counter), without requiring
explicit recurrent state.

\bibliography{references}
%% =====================================================================

\end{document}